\documentclass[runningheads]{llncs} 

\title{Case Study: Disclosure of Indirect Device Fingerprinting in Privacy Policies}

\author{Julissa Milligan \and
Sarah Scheffler\thanks{Supported by a Clare Boothe Luce Graduate Research Fellowship and a Google PhD Fellowship.} \and
Andrew Sellars \and
Trishita Tiwari \and
Ari Trachtenberg \and
Mayank Varia}

\institute{Boston University\thanks{This material is based on work supported by the National Science Foundation under Grants CNS-1414119, CCF-1563753, and CNS-1915763.}}

\usepackage{cite}
\usepackage{graphicx}
\usepackage{hyperref}

\begin{document}

\maketitle
\thispagestyle{empty}
\pagestyle{empty}

\begin{abstract}
Recent developments in online tracking make it harder for individuals to detect and block trackers.  This is especially true for device fingerprinting techniques that websites use to identify and track individual devices.  Direct trackers -- those that directly ask the device for identifying information -- can often be blocked with browser configurations or other simple techniques.  However, some sites have shifted to indirect tracking methods, which attempt to uniquely identify a device by asking the browser to perform a seemingly-unrelated task.  One type of indirect tracking known as \emph{Canvas fingerprinting} causes the browser to render a graphic recording rendering statistics as a unique identifier.  Even experts find it challenging to discern some indirect fingerprinting methods. 
%
In this work,
we aim to observe how indirect device fingerprinting methods are disclosed in privacy policies, and consider whether the disclosures are sufficient to enable website visitors to block the tracking methods.  We compare these disclosures to the disclosure of direct fingerprinting methods on the same websites.

Our case study analyzes one indirect fingerprinting technique, \emph{Canvas fingerprinting}. We use an existing automated detector of this fingerprinting technique to conservatively detect its use on Alexa Top 500 websites that cater to United States consumers, 
and we examine the privacy policies of the resulting 28 websites.
%
Disclosures of indirect fingerprinting vary in specificity.  None described the specific methods with enough granularity to know the website used Canvas fingerprinting.  Conversely, many sites did provide enough detail about usage of direct fingerprinting methods to allow a website visitor to reliably detect and block those techniques. 

We conclude that indirect fingerprinting methods are often technically difficult to detect, and are not identified with specificity in legal privacy notices.  This makes indirect fingerprinting more difficult to block, and therefore risks disturbing the tentative armistice between individuals and websites currently in place for direct fingerprinting.  This paper illustrates differences in fingerprinting approaches, and explains why technologists, technology lawyers, and policymakers need to appreciate the challenges of indirect fingerprinting. 

\end{abstract}


\section{Introduction}
Companies employ a variety of ``fingerprinting'' techniques to track consumers' identities online. These fingerprints are used to identify -- or at least significantly narrow the range of possibilities for -- repeated visits to a site by the same device or individual, or within the same location or web browsing session. In this work, we focus on device fingerprinting over the web: code on the server of a website (rather than in email or dedicated applications) that seeks to uniquely identify each consumer device that visits the site.  Device fingerprinting can often identify a single device in a manner that persists across browsing sessions, that is throughout usage at different physical and virtual locations and among different people. We provide a brief summary of direct and indirect methods for device fingerprinting in the next section, and we refer interested readers to the following surveys for more detail \cite{alaca2016device,acar2014web,nikiforakis2013cookieless,eckersley2010unique}.

From a purely technological viewpoint, the current evolution of device fingerprinting is akin to a cat-and-mouse game. Companies seek increasingly detailed data and collection techniques to increase their confidence that they have identified a similar device across different visits. Individuals can respond by declining to visit certain sites or using blocking software that seeks to prevent those companies from obtaining information that could be used to identify the visitors.  Companies can then refuse to serve their site to individuals using blockers, and so on. This cat-and-mouse game plays out across many websites.

Anecdotally, much of the web seems to have settled upon a kind of d\'etente at the intersection of technology and policy: companies disclose that they track devices and use reasonably transparent (or at least detectable) ``direct'' tracking technologies, and the small percentage of consumers who do object to such tracking use technological tools such as ad blockers to inhibit tracking.  Many companies nevertheless welcome these tracking-inhibiting visitors on their sites. 

``Indirect'' or ``inference-based'' fingerprinting works differently, using methods that serve a purpose unrelated to tracking, like the HTML5 Canvas API, to develop a unique device identifier.  Because these techniques are typically dual-use -- that is, they can fingerprint a user or alternatively perform different user-friendly function on the website -- it is more challenging to detect whether they are used as trackers.  Indeed, there are few public tools that can detect or block indirect fingerprinting, and these tools might themselves be detected by websites and used to fingerprint \cite{multilogin}.

A privacy-aware user might therefore turn to a company's privacy policy to determine whether a site uses indirect fingerprinting techniques.  One purpose of such privacy policies is to tell consumers what type of data the site collects.  A technologically savvy individual could, in theory, review the disclosed fingerprinting methods and design a technical response that meets her comfort level, which the website could accept or reject.  In practice, many sites strike a balance between disclosure and readability, and thus they purposely avoid some technical details about how data is collected.  But if a privacy policy does not contain technical details about which indirect tracking methods it uses, creating a technical block against indirect fingerprinting becomes much more difficult.

In this paper, we examine changes in tracking technology over the past half-decade, and study how websites explain Canvas fingerprinting in their privacy policies.  We illustrate the differences between direct and indirect fingerprinting.  We analyze the disclosure of indirect fingerprinting in privacy policies to consider how these techniques may destabilize the delicate direct-fingerprinting truce between websites and visitors.  Finally, we consider whether these differences are important to potential technical and legal responses.

\section{Device Fingerprinting}

\subsection{Direct Fingerprinting}
One common way to identify a device is to directly ask the device for identifying information. For example, websites can use one of several Application Programming Interface (API) calls to elicit a client browser to send device information, such as its operating system or Internet Protocol (IP) address, or to store identifying information locally for future use (e.g., cookies). Widely-deployed techniques that websites use in this manner include:

\begin{itemize}
\item Collecting header information transmitted through the standard HTTP exchange, and using the uniqueness of that data to develop a distinguishable profile;
\item Embedding discreet objects (web beacons, tracking pixels, or clear GIFs) within a common third-party website, which can be used to track access patterns; or
\item Placing a cookie on the site, either directly or through a third-party tracker.
\end{itemize}
These techniques enable web servers both to personalize web services (e.g., for language or region) and to fingerprint the device, often simultaneously or in an intertwined manner.

Direct fingerprinting methods are easy to detect, understand, and reset. When used, direct fingerprinting techniques are typically simple to detect because they operate similarly on all websites and rarely obfuscate their true intention. Novice users can use a web browser's built-in features or download a plugin to identify their use. Additionally, expert consumers can examine a log of all interactions between the web server and the client browser to understand precisely the type of information obtained by the server and (unless it is encrypted at rest or in transit by the server) read its contents. Once detected, many direct fingerprinting methods have ``reset'' mechanisms that decouple future visits to a website from previous ones; for instance, deleting a cookie with a unique identifier can have this effect.

Individuals can also proactively block unwanted direct tracking by appropriately configuring their client. For example, if a visitor does not want a web server to learn her device's operating system, she can configure her browser to block or falsify the responses to common requests that reveal her operating system, including the \texttt{User-Agent} field of the HTTP header, and the \texttt{navigator.userAgent} and \texttt{navigator.platform} properties of the \texttt{navigator} object within JavaScript.  Indeed, modern web browsers give users the ability to block or control cookie storage, and third-party tools like Adblock Plus and Privacy Badger allow users to selectively block connections and cookies from websites that are known to serve ads, web beacons, and other direct trackers. These blocks can be persistent, in that blocking one direct request for specific information will block all future requests for the same information. In this way, a user is empowered to transform the binary choice presented by accessing a website -- use the website with these tracking features, or do not use the website at all -- into a negotiated environment. She can use the website and control some of the information that the website obtains about her. 

\subsection{Inference-Based, or Indirect, Fingerprinting}
In contrast, \emph{inference-based device fingerprinting} is a newer set of tracking techniques that use different tools to achieve the same goal. Rather than directly querying the browser about its localizing features, the server will instead instruct the browser to perform a seemingly-unrelated computing task such as rendering text, audio, a picture, or an animation. Different devices with different configurations, installed libraries, and hardware will perform the task in slightly different ways, and these subtle differences can be measured and summarized by the server, creating a fingerprint for the device. Many techniques for web servers to conduct indirect device fingerprinting have arisen over the last decade, including:

\begin{itemize}
\item Measurements of JavaScript performance \cite{mowery2011fingerprinting} and conformance \cite{mulazzani2013fast},
\item Font enumeration \cite{fifield2015fingerprinting},
\item Graphics Processing Unit (GPU) behavior measurement~\cite{cao2017browser},
\item VRAM detection \cite[p.~5]{alaca2016device},
\item Sensor data access \cite{das2018web}, and
\item Techniques that use HTML5 APIs \cite{nakibly2015hardware}.
\end{itemize}

\begin{figure}
\centering
\includegraphics[width=0.5\textwidth]{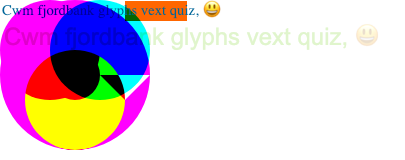}
\caption{Graphic sent to clients to draw in Canvas fingerprinting script within open-source fingerprinting script \url{https://github.com/Valve/fingerprintjs2}, commit 563dbde.  The way in which the graphic is drawn can be used to fingerprint the client's device.} \label{fig:fingerprintjs2-image}
\end{figure}

One of these HTML5 APIs is known as the Canvas API. Normally used for rendering graphics or video on a screen~\cite{canvasapi}, the Canvas can also be used to fingerprint devices \cite{mowery2012pixel} by instructing the client device to render some text or gradients in the client browser, and then reading back the exact pixel data of the image rendered by the browser.  The image used in a popular open-source fingerprinting script  \texttt{fingerprintjs2} is shown in Figure \ref{fig:fingerprintjs2-image}.  The resulting fingerprints are highly effective because they are both highly distinguishing (a large number of machines yield different renderings) and highly stable (the same machine repeatedly yields the same result). The team behind the anonymity-seeking Tor Browser has called Canvas fingerprinting ``the single largest fingerprinting threat browsers face today''~\cite{torbrowser}, aside from plugins such as Flash. Indeed, Englehardt and Narayanan \cite{openWPM} also performed an extensive study in 2016 on the top 1 million websites, and found that 14,371 of those websites employed Canvas fingerprinting. Of these, they found that 98.2\% of the Canvas fingerprinting scripts came from third party websites from around 400 domains.

In this study, we focus on Canvas fingerprinting because it is a reasonably detectable indirect fingerprinting method that has been well-studied over the last six years, and there are consequently well-justified heuristics provided by Acar et al.~\cite{acar2014web} and Englehardt and Narayanan~\cite{openWPM} that distinguish fingerprinting uses of Canvas from non-fingerprinting uses.

Unlike direct tracking techniques, there are relatively few tools available to detect or understand Canvas fingerprinting. To analyze whether a Canvas query is being used at all requires a deeper inspection of the interchange between server and client than inspecting one's own HTTP header, identifying a cookie request, or observing third-party websites calls in the chain of HTTP requests made when loading a website. Even if an individual does observe the use of a Canvas query, it is hard to know whether it is used to fingerprint a device or for some other legitimate graphical purpose. A Canvas API call could use the information from the web client to create a unique fingerprint, to check whether the client rendered an image correctly, or to determine how to properly organize information on the client's screen. To distinguish between these options, the user would need to predict how the server is (or will be) processing the resulting data. 

Furthermore, tools that allow consumers to automatically mask, reset, or block Canvas queries are not as well-known as their counterparts to block direct tracking. For example, Adblock Plus, a common ad-blocker, has 11 million average daily users on Firefox. In contrast, no public tools to effectively block the fingerprinting existed in 2014~\cite{acar2014web}, and only a few browser add-ons/extensions are available today. The most popular Firefox add-on for blocking Canvas fingerprinting has only about 46,000 average daily users as of August 2019. 

These tools also may not persistently block the fingerprinting, because the precise way that a website configures a Canvas query to fingerprint a device may change over time.  Common tools to block direct fingerprinting tend to target fixed web content and objects -- most commonly HTTP header information, cookies, and connections to third-party websites. With Canvas queries, however, the fingerprinting scripts could easily be changed, obfuscated, or combined with commonly used scripts. This could make it more difficult to detect or block fingerprinting scripts without breaking the functionality of many websites.  In our experiments, we were able to easily detect Canvas fingerprinting websites that use minor variations of five dedicated fingerprinting scripts. It is possible we did not detect existing obfuscated fingerprinting techniques.

\section{What Do Privacy Policies Say About Fingerprinting?}

We studied the ways tracking techniques are discussed in privacy disclosures to better understand how users might learn about direct versus indirect tracking. Specifically, we examined 28 privacy policies of websites that appear to use Canvas queries for fingerprinting purposes. In the United States, the collection of consumer browsing data on commercial websites is generally regulated by the Federal Trade Commission, which polices unfair and deceptive acts or practices in or affecting commerce.%
\footnote{15 U.S.C. \S 45(a)(1),(n).}
When a company is not operating in a sector that is specifically regulated by another statute (e.g., healthcare or finance), the primary obligation on companies is to provide accurate information about how the site collects and discloses user data so that a reasonable consumer has meaningful choice about whether to submit to those practices.%
\footnote{See, e.g., \emph{In re Liberty Fin. Cos., FTC No. C-3891}.}

A privacy policy for a company could reveal information about fingerprinting by disclosing the specific information it gathers, or by discussing the techniques it uses to obtain information.  Many policies use some combination of these disclosures.  With respect to fingerprinting, a policy could specifically explain whether and/or how the company uses fingerprinting technology, or it could discuss what data it collects, leaving the inference about which fingerprinting technique it uses to the consumer.  Of course, it could also disclose both technology and data.  We examine the level of specificity of these privacy policy disclosures below. These disclosures provide a useful base from which to consider further whether
indirect fingerprinting methods, such as Canvas, are well-known or widely-publicized techniques. 

\subsection{Methodology}

To find instances of Canvas fingerprinting on popular websites, we conducted two web crawls of the Alexa top 500 websites using the code accompanying Acar et al.'s 2014 study \cite{acar2014web}. Run 1 ran on January 15, 2019 from 2:19am to 1:55pm EST and found Canvas fingerprinting on 40 out of 470 successful connections.  Run 2 ran from January 15 at 4:22pm EST to January 16 at 3:48pm EST and found Canvas fingerprinting on 42 out of 484 successful connections.  In total, across both runs, we found 49 unique websites that had fingerprinting scripts.

Because our expertise is focused on privacy law within the United States,
we manually inspected these 49 webpages and filtered them based on the following two criteria.
\begin{enumerate}
 \item The website's main page is written in English (30 of 49 websites).
 \item The website and its associated privacy policy are written for an audience of U.S.~consumers. We discarded 1 site with the country-specific domain \texttt{.co.uk} and 1 site for an Indian bank lacking a physical presence in the U.S.
\end{enumerate}
We manually reviewed the privacy policies of the remaining 28 websites%
\footnote{
These sites are
americanexpress.com,
aol.com,
asos.com,
bestbuy.com,
cambridge.org,
capitalone.com,
cnet.com,
coinmarketcap.com,
dell.com,
drugs.com,
fiverr.com,
forbes.com,
foxnews.com,
homedepot.com,
ikea.com,
livescore.com,
msn.com,
nike.com,
shutterstock.com,
slickdeals.net,
sonyliv.com,
speedtest.net,
thesaurus.com,
udemy.com,
upwork.com,
weather.com,
yelp.com,
and
zillow.com.}
between January-July 2019 to determine which disclosures in their policies are intended to inform consumers about their use of Canvas fingerprinting.
At least one legal scholar and one technologist on our team read each privacy policy and identified all statements that explicitly or implicitly refer to device fingerprinting.

\subsection{Results}

All 28 privacy policies share two features in common: they all indicate that the site uniquely identifies individual devices, and none of them state specifically that the site uses Canvas fingerprinting.
Otherwise, the 28 privacy policies vary substantially in the specificity of disclosures regarding tracking techniques.
We identified three categories of privacy policy disclosures of Canvas fingerprinting:
(1) broad, technology-agnostic language, (2) disclosure of specific direct fingerprinting techniques but not indirect techniques, and (3) specific mention of collecting ``device fingerprints'' as distinct from other device-specific identifiers.

One set of privacy policies uses very broad and technology-agnostic language that covers the capture of granular, device fingerprinting data, but provides little guidance on specific techniques or types of information captured. For example, the privacy policy of the website CNET (which is part of CBS, and uses CBS' general Privacy Policy) states that it collects ``[u]nique identifiers and connection information'' of various sorts~\cite{CBS}. 
Other sites simply note that they collect ``information about your use of the Website, and/or mobile application''~\cite{coinmarketcap}.

A second group of policies provide more detailed information about how a site collects data using older tracking technologies, and less specific information about the types of data they collect through fingerprinting.  The majority of privacy policies investigated fell into this category.
Yelp, for example, states that it may store data ``such as your browser type, type of computer or mobile device, browser language, IP address, WiFi information such as SSID, mobile carrier, phone number, unique device identifier, advertising identifier, location (including geolocation, beacon-based location, and GPS location), and requested and referring URLs.''~\cite{Yelp}  This list appears to specifically call out the data typically obtained from direct tracking techniques like cookies or HTTP header information. The list is non-exhaustive, and fingerprinting appears to be covered under the open-ended ``unique device identifier." That said, the specific disclosures appear more tailored to older tracking techniques than to fingerprinting. 
The Fox News Privacy Policy states that ``[i]f you access the Fox News Services from a mobile or other device, we may collect a unique device identifier assigned to that device, geolocation data (including your precise location), or other transactional information for that device,'' which reads broadly enough to embrace Canvas data~\cite{FoxNews}. 
Forbes lists its collection by identification technique, specifically stating that it collects data through ``cookies,'' ``web beacons,'' and ``log files.''  It also acknowledges that Forbes may retain any ``information automatically collected about your usage of the Site'' \cite{Forbes}. 
That open-ended disclosure may well encompass fingerprinting techniques, but the policy does not expressly identify Canvas or other indirect fingerprinting technologies in the same way it identifies some direct tracking techniques.

We also identified one policy that specifically states the website uses device fingerprinting, but it does not call out the fingerprinting technique that it uses -- Canvas fingerprinting -- or explain what data it collects using Canvas fingerprinting.  Udemy's privacy policy lists ``device or browser fingerprints'' along with other tracking methods it uses \cite{udemy}.  Describing device fingerprinting specifically \emph{in addition} to all the other information collected suggests that the website is indirectly obtaining a fingerprint by measuring the device's responses (e.g. to Canvas queries), as opposed to simply using direct queries, even of device-specific IDs.

\subsection{Observations}

In short, most privacy policies generally describe \emph{what} information is collected rather than providing details on the \emph{method} of collection. 
Those that do list specific tracking technologies tend to omit fingerprinting. And the site that listed fingerprinting did not state which methods of fingerprinting it deployed.
This may suggest that the public and perhaps the lawyers who draft privacy policies may not yet be aware of more recent indirect fingerprinting techniques. 

Of note, some of the privacy policies may be written to comply with the European Union's General Data Protection Regulation (GDPR) \cite{gdpr}. If the GDPR requires attorneys to more deeply understand and explain the website's tracking practices at a technical level, we would expect to see greater clarity in the way in which fingerprinting, or the data collected thereby, is described in privacy policies. 
Companies may seek to harmonize policies across jurisdictions, and this increased understanding and explanation could
benefit people from non-EU member nations (like the United States) as well. However, that shift did not jump out in the policies we reviewed at this point in time. 

More to the point, even when a policy states that the website uses fingerprinting, it may be difficult to identify which indirect fingerprinting methods the website uses. All companies in our review indicated that they collect information that could be used to identify a device -- putting consumers on notice that the site likely correlates user browsing behavior over time.
However, the policies we reviewed did not provide enough information about how data is collected to allow even savvy individuals to access the website while proactively blocking such collection or retrospectively resetting an identifier held by the website.

\subsection{Consumer responses}

For direct fingerprinting methods, this gap appears to be surmountable.  A consumer wishing to hide his browser type may alter her \texttt{User-Agent} and the relevant fields of the JavaScript \texttt{navigator} object, since these are the relevant methods a website might use to directly query this information.  She need not know the details of the server's code, since there are limited methods to query this information directly. 

However, for indirect fingerprinting methods, even a consumer who is both legally and technologically savvy finds little recourse to tracking from either of her areas of expertise. The individual must first know that a functionality seemingly unrelated to fingerprinting (like Canvas) can be used to fingerprint.  She must then inspect each use of the functionality and infer whether the purpose is related to fingerprinting or not.
This process could involve an inspection of client-side code (e.g. the JavaScript code generating the image in Figure~\ref{fig:fingerprintjs2-image}),
or heuristics that make an educated guess as to how the server-side code will process the information it receives.
The website's privacy policy does not provide her with any guidance on how to perform any of the above steps.
Ultimately, to use the website without being fingerprinted, she must choose between being cautious by overzealously blocking useful website components, or being overconfident that she has blocked all the tracking that is personally objectionable and missing an indirect fingerprinting method.

\section{Indirect Fingerprinting Shifts the Balance Between Individuals and Websites}

Indirect fingerprinting can be technologically difficult to identify, and may not be specifically identified in privacy policies.  These properties make indirect fingerprinting different than direct methods, where a rough armistice has evolved between consumers and websites.

\subsection{Disturbing a delicate armistice}

Direct fingerprinting techniques are often identifiable technologically. They can also be inferred from the disclosures made in the privacy policy, either because they are explicitly mentioned or because the policy discloses that it collects data typically obtained using direct fingerprinting.
Once detected, tools are available (although not universally adopted) to thwart such attempts at fingerprinting.  Many of these tools have been adopted by enough consumers that many  websites complain about the use of these techniques and their effect on advertising revenue. 
Websites have a technological countermeasure of their own: they can detect and deny access to consumers who deploy ad-blocking technologies. While some websites use these anti-blocking tools, most websites opt to accept privacy-aware consumers who block direct fingerprinting.
Put simply: we think the ad blocking arms race has reached a de facto armistice in which privacy-aware consumers can choose whether to block direct fingerprinting, and websites can choose whether to accept consumers who opt to block.

However, indirect fingerprinting disrupts this armistice.  Indirect fingerprinting techniques are harder to technically detect and block, and easier for the company to change and obfuscate. These techniques can serve a functional end beyond fingerprinting a device, so it is harder to predict whether the website is using the relevant technique to persistently identify the user or for some other purpose.
Indirect fingerprinting methods are not yet well-known to the general public, and 
privacy policies do not specifically describe indirect device fingerprinting methods -- even policies that do describe direct methods.
In short, consumers do not have an accessible way to learn a website's indirect fingerprinting practices.

In the short term, individuals who are unaware of indirect fingerprinting techniques may mistakenly believe that tools like Adblock Plus and Privacy Badger are sufficient to prevent tracking. 
In the longer run, individuals (or the developers of privacy tools they use) will need to be aware of evolving ways to fingerprint devices indirectly using an existing API call, and develop methods to detect and block each of these new methods as they are invented.

\subsection{A path forward}

The authors of this article have different views about the best path forward and whether any alternative is better than the current state of affairs.
The spectrum of options ranges from 
maintaining the status quo
while calling attention to these new techniques, to removing consumer choice entirely and imposing privacy defaults by law.

The area between these options illustrates multiple tradeoffs.
For example, changing FTC guidance to instruct websites to list all tracking practices would 
increase public information about indirect tracking but would
make privacy policies longer, more cumbersome, and less comprehensible to non-technical readers.
Requiring companies by law to provide legal or technological means to block and/or reset individual identifiers would increase consumer choice
but would put new burdens on companies and could adversely impact competition.

It may be possible to use a combination of tools to increase consumer choice while minimizing other costs.  For instance, a social campaign to heighten public awareness about device fingerprinting might spur consumers to choose sites that avoid fingerprinting.  Alternatively, larger liability for breaches of unique device identifiers might lead websites to reconsider the value of gathering them.

This work brought together authors with different fundamental perspectives
about the appropriate role for law and regulation in the context of privacy and consumer protection.
Despite our differences, we reached consensus on the technological and legal state of affairs
and improved our understanding of the tradeoffs.
We hope this paper lays the foundation for additional conversations and improves the quality of the debate.



\bibliographystyle{abbrv}
\bibliography{main}

\end{document}